\documentclass[aps,prb,reprint,superscriptaddress]{revtex4-2}
\usepackage{amsmath}
\usepackage{amssymb}
\usepackage{graphicx}
\usepackage{hyperref}
\usepackage{color,xcolor}
\begin{document}
\title{Strong tuning of magnetism and electronic structure by spin orientation}
\author{Yakui Weng}
\author{Xing'ao Li}
\affiliation{School of Science, Nanjing University of Posts and Telecommunications, Nanjing 210023, China}
\author{Shuai Dong}
\email{Corresponding author. Email: sdong@seu.edu.cn}
\affiliation{School of Physics, Southeast University, Nanjing 211189, China}
\date{\today}

\begin{abstract}
To efficiently manipulate magnetism is a key physical issue for modern condensed matter physics, which is also crucial for magnetic functional applications. Most previous relevant studies rely on the tuning of spin texture, while the spin orientation is often negligible. As an exception, spin-orbit coupled $J_{\rm eff}$ states of $4d$/$5d$ electrons provide an ideal platform for emergent quantum effects. However, many expectations have not been realized due to the complexities of real materials. Thus the pursuit for more ideal $J_{\rm eff}$ states remains ongoing. Here a near-ideal $J_{\rm eff}$=$3/2$ Mott insulating phase is predicted in the family of hexachloro niobates, which avoid some common drawbacks of perovskite oxides. The local magnetic moment is nearly compensated between spin and orbital components, rendering exotic recessive magnetism. More interestingly, the electronic structure and magnetism can be strongly tuned by rotating spin axis, which is rare but crucial for spintronic applications.
\end{abstract}
\maketitle

\begin{figure}
\centering
\includegraphics[width=0.46\textwidth]{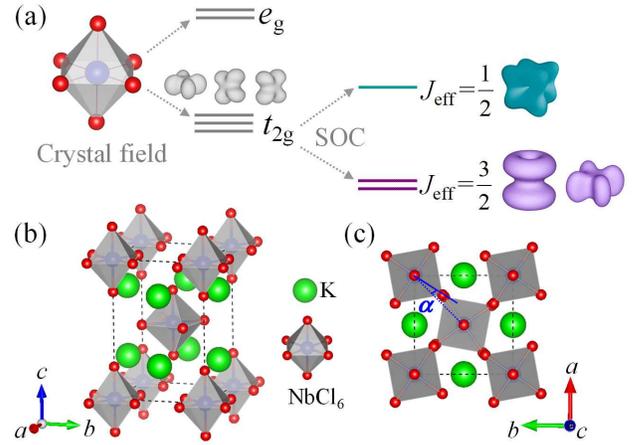}
\caption{(a) Splitting of $d$ orbitals by octahedral crystal field and SOC. The crystal field leads to three low-lying $t_{\rm 2g}$ levels and two higher energy $e_{\rm g}$ doublets. The SOC further splits the $t_{\rm 2g}$ triplets to two $J_{\rm eff}$ states: $1/2$ and $3/2$ ones. The corresponding orbital shapes are presented. Further considering the spin degeneracy, all these states are duplicated, i.e. two $J_{\rm eff}=1/2$ and four $J_{\rm eff}=3/2$ ones. (b-c) Crystal structure of K$_2$NbCl$_6$ (space group No. $128$ $P4/mnc$). (b) Side view; (c) Top view. $\alpha$ denotes the NbCl$_6$ octahedral rotation angle with respect to the [$110$] axis. Green: K; blue: Nb; red: Cl.}
\label{F1}
\end{figure}

\textit{Introduction}. The spin-orbit entangled quantum states open new frontiers of condensed matter, which can manifest novel physics such as topological bands, quantum spin liquid, as well as unconventional superconductivity \cite{Krempa:Arcmp,Rau:Arcmp16,Cao:Bok,Pesin:Np10}. For example, the $J_{\rm eff}$=$1/2$ Mott insulating state as proposed in Sr$_2$IrO$_4$ \cite{Kim:Prl08}, might be closely related to high-temperature superconductivity \cite{Wang:Prl11,Watanabe:Prl13,Kim:Sci14,Yan:Prx15,Kim:Np16} and large anisotropic magnetoresisitive \cite{Lu:Afm18,Wang:Nc19}, which is a collaborative result of strong SOC and moderate Hubbard correlation. Although the $J_{\rm eff}$ scenario is quite elegant in the atomic limit [as sketched in Fig.~\ref{F1}(a)], real materials always deviate from the ideal limit more or less, which makes many novel expectations become unrealistic. For example, the Kitaev spin liquid was expected for $J_{\rm eff}$=$1/2$ state on honeycomb lattice \cite{Kitaev:Ar06,Jackeli:Prl09,Takagi:Nrp19}, e.g. Na$_2$IrO$_3$ \cite{Chaloupka:Prl10,Chun:Np15}, $\alpha$-Li$_2$IrO$_3$ \cite{Manni:Prb14,Caog:Prb13}, and $\alpha$-RuCl$_3$ \cite{Banerjee:Nm16,Zheng:Prl17,Baek:Prl17}, but has not been realized yet in these materials. Even for the prototype $J_{\rm eff}$=$1/2$ Mott state in Sr$_2$IrO$_4$, the calculated magnetic moments of Ir$^{4+}$ are $0.10$ $\mu_{\rm B}$ from spin and $0.26$ $\mu_{\rm B}$ from orbital contributions, far from the expected $1/3$ $\mu_{\rm B}$ from spin and $2/3$ $\mu_{\rm B}$ from orbital \cite{Kim:Prl08}. Such deviations can be due to other interactions beyond the SOC and Hubbard correlation. For example, the $Q_3$ mode of Jahn-Teller distortion associated with the layered structure of Sr$_2$IrO$_4$ breaks the degeneration between $d_{xy}$ and $d_{yz}$/$d_{xz}$. And the highly extending $5d$ electron clouds lead to wide $5d$ bands, which heavily hybridize with oxygen's $2p$ orbitals.

Searching for new materials to host the ideal $J_{\rm eff}$ states is only the first step, the more important issue is to manipulate these states, especially to achieve some valuable functions. Although a recent experiment on Sr$_2$IrO$_4$ found that the rotation of magnetic axis can tune the transport properties in this non-ideal $J_{\rm eff}$=$1/2$ system \cite{Lu:Afm18}, more significant effects should be expected for the $J_{\rm eff}$=$3/2$ case considering the anisotropy of orbital shapes as sketched in Fig.~\ref{F1}(a). In fact, the $J_{\rm eff}$=$3/2$ state is more interesting. In the ideal limit, its net magnetic moment is completely cancelled between orbit and spin components, leading to an exotic recessive magnetism.

However, the $J_{\rm eff}$=$3/2$ state is much less studied. Till now, there are three candidate families of $J_{\rm eff}$=$3/2$ state. One branch is the double perovskite family, e.g., Ba$_2$YMoO$_6$ \cite{Vries:Prl10,Aharen:Prb10,Carlo:Prb11}, Ba$_2$NaOsO$_6$ \cite{Erickson:Prl07,Gangopadhyay:Prb15}, Sr$_2$MgReO$_6$ \cite{Wiebe:Prb03}, and Ba$_2$MgReO$_6$ \cite{Hirai:Jpsj}. In these systems, each heavy $B'$ ion with strong SOC is isolated by six nearest-neighbor nonmagnetic $B$ ions, and thus their $d$-orbital bands are largely narrowed, closer to the atomic limit. Also the non-layered structure suppresses the Jahn-Teller $Q_3$ mode distortion. However, the strong $d$-$p$ hybridization remains 
\cite{Xiang:Prb07,Gangopadhyay:Prb15}. Another predicted family is the lacunar spinel Ga$M_4X_8$ ($M$=Nb, Mo, Ta, W and $X$=S, Se, Te) \cite{Kim:Nc14}, while one member GaTa$_4$Se$_8$ has been experimentally confirmed to host the molecular $J_{\rm eff}$=$3/2$ state \cite{Jeong:Nc17}. Very recently, $M_2$TaCl$_6$ ($M$=K, Rb, Cs) was studied experimentally \cite{Ishikawa:Prb19}, which hosts the $5d^1$ $J_{\rm eff}$=$3/2$ state. However, it remains unclear whether $M_2$TaCl$_6$ are SOC Mott insulators.

In this article, the $J_{\rm eff}$=$3/2$ state in hexachloro niobate K$_2$NbCl$_6$ will be studied by density functional theory (DFT). Comparing with their sister compound $M_2$TaCl$_6$, the relative stronger Hubbard correlation and spatial localization of $4d$ orbitals is advantageous for the SOC Mottness. The $J_{\rm eff}$=$3/2$ Mott insulating phase is unambiguously revealed, which is close to the ideal $J_{\rm eff}$=$3/2$ atomic limit. More importantly, one of its novel physical properties, i.e., the strong tuning of magnetism and electronic structure by spin orientation, is demonstrated, which is physically interesting and valuable for appliations.

\textit{Model and method}. As shown in Fig.~\ref{F1}(b-c), the crystal structure of K$_2$NbCl$_6$ is tetragonal when temperature is below $282$ K \cite{Henke:Zk}. The slight orthorhombic distortion at low temperature \cite{Henke:Zk} will not affect our conclusion, as demonstrated in Supplemental Materials (SM) \cite{Supp} (see, also, references \cite{Soderlind:Prb10,Heyd:Jcp,Heyd:Jcp06,WangY:Prb19,LiuP:Prm20,He:Prb12,Henke:Zk} therein). Considering the nearly isolated NbCl$_6$ octahedra (i.e. the absence of Nb-Cl-Nb bonding), the electron hopping between Nb ions is significantly suppressed and hence the $4d$ bands near the Fermi level are expected to be narrow. In addition, since Nb$^{4+}$ has only one $4d$ electron, the singlet $J_{\rm eff}$=$3/2$ Mott-insulating phase is expectable.

DFT calculations were performed using the projector augmented wave (PAW) pseudopotentials as implemented in the Vienna $ab$ $initio$ simulation package (VASP) code \cite{Blochl:Prb2,Kresse:Prb99}. The revised Perdew-Burke-Ernzerhof for solids (PBEsol) functional and the generalized gradient approximation (GGA) method are adopted to describe the crystalline structure and electron correlation \cite{Perdew:Prl08}. The PBE and hybrid functional calculations based on the Heyd-Scuseria-Ernzerhof (HSE06) exchange are also tested for comparison \cite{Heyd:Jcp,Heyd:Jcp06}, which do not alter the physical conclusion \cite{Supp}.

The cutoff energy of plane-wave is $400$ eV and the $11\times11\times7$ Monkhorst-Pack \textit{k}-point mesh is centered at $\varGamma$ point. Starting from the experimental structure, the lattice constants and atomic positions are fully relaxed until the force on each atom are converged to less than $0.01$ eV/{\AA}. For the spin polarized LSDA+$U$(+SOC) (LDAUTYPE=$2$) calculation, the value of $U_{\rm eff}$=$U-J$ (Dudarev approach) \cite{Dudarev:Prb} is tuned from $0$ to $2$ eV. In addition, the plain LDA+$U$(+SOC) calculation (LDAUTYPE=$4$, i.e. no LSDA exchange splitting) is also done to verify the result, as shown in SM \cite{Supp}. Ferromagnetic spin order is adopted, which is the most stable state when SOC is included \cite{Supp}.

\begin{figure}
\centering
\includegraphics[width=0.46\textwidth]{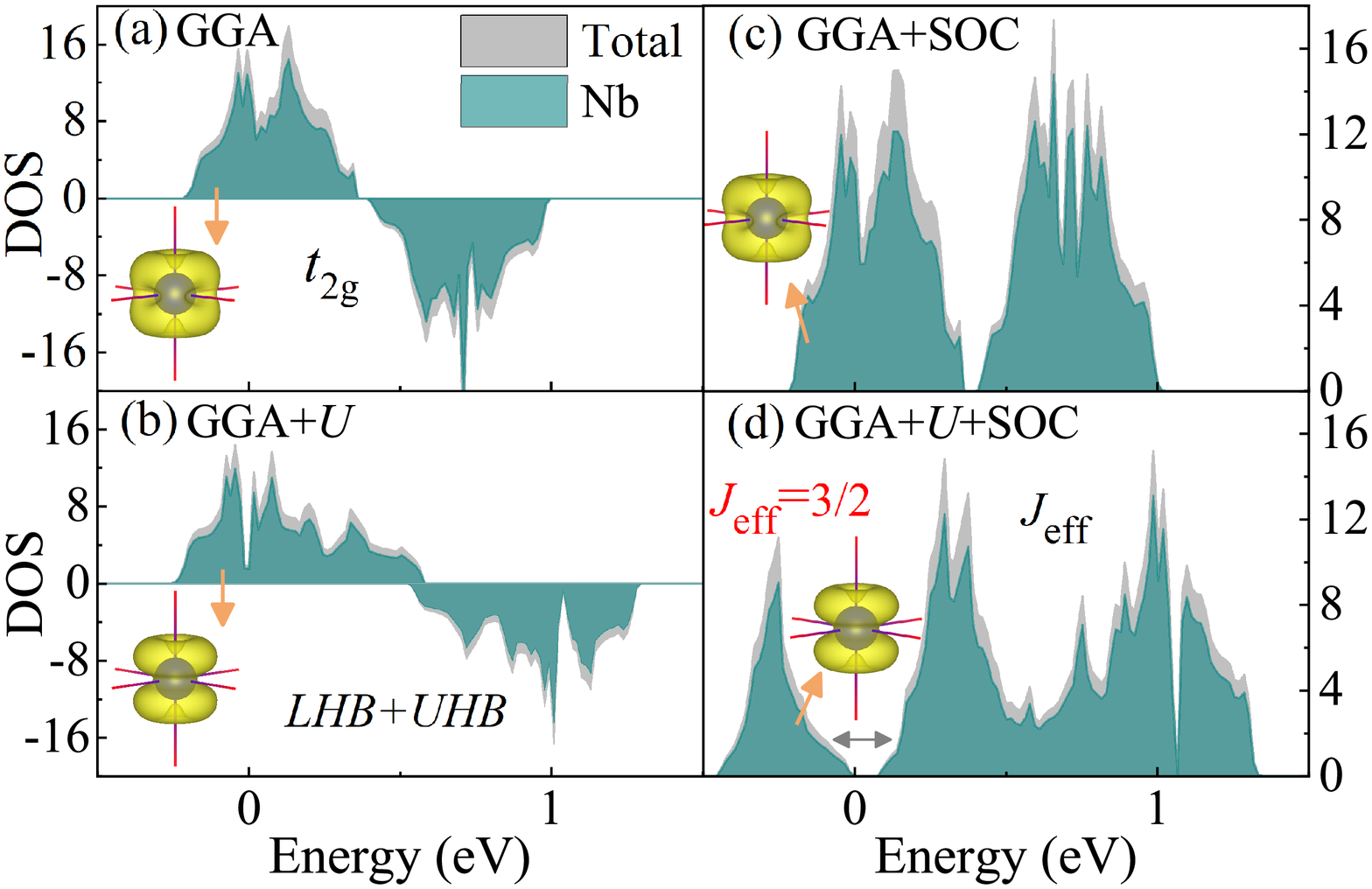}
\includegraphics[width=0.46\textwidth]{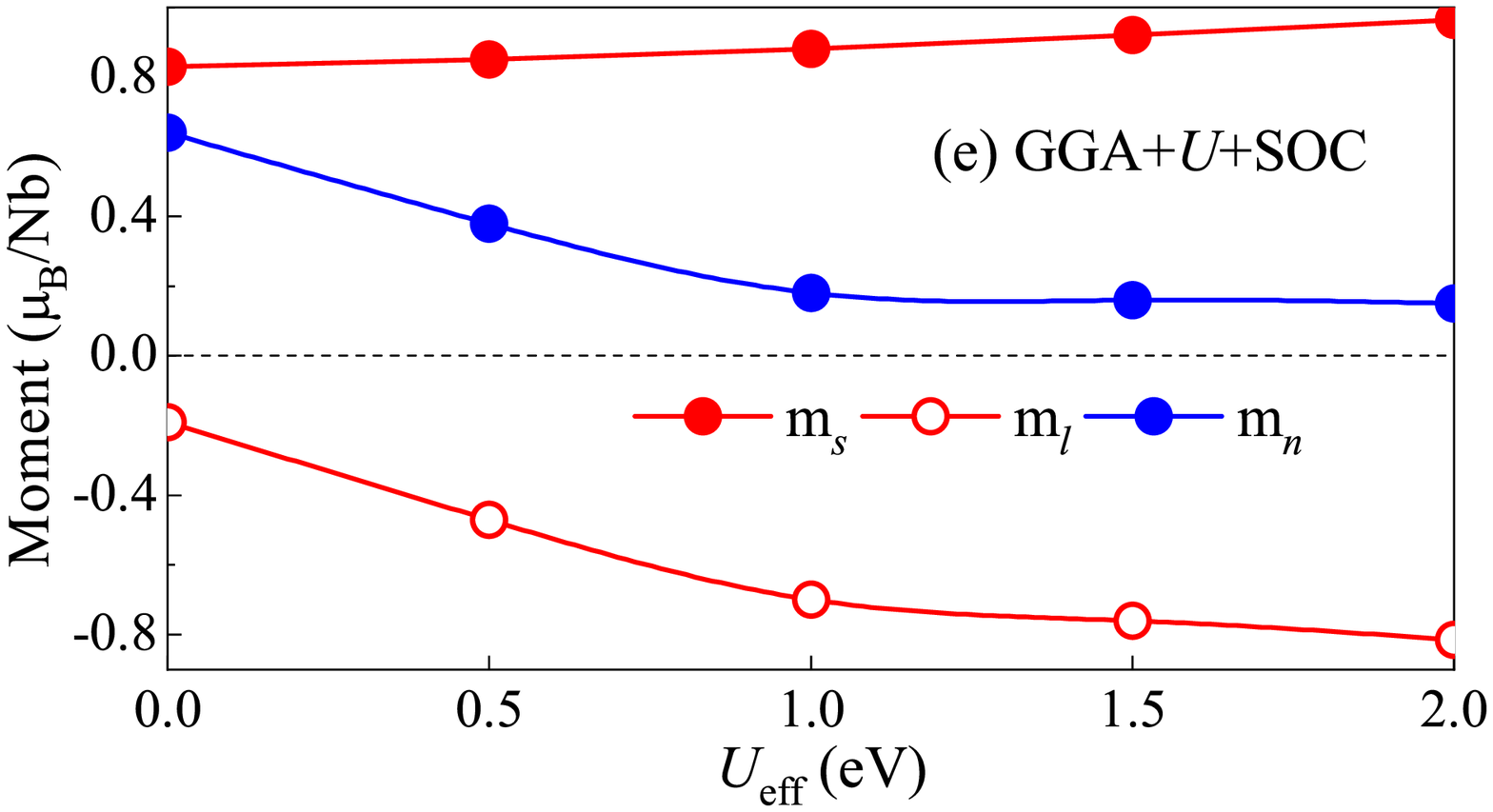}
\caption{Total density of states (DOS) (gray) and atom-projected DOS (PDOS) (cyan) of K$_2$NbCl$_6$ near the Fermi level. (a) Calculated using generalized gradient approximation (GGA); (b) GGA+$U$; (c) GGA+SOC; (d) GGA+$U$+SOC. The Fermi energy is positioned at zero. In (b) and (d), $U_{\rm eff}$=$1$ eV is applied on Nb's $4d$ orbitals. Inset: the electron cloud of valence bands near the Fermi level. (e) The local spin ($\textbf{m}_s$), orbital ($\textbf{m}_l$) and net magnetic moments ($\textbf{m}_n$=$\textbf{m}_s$+$\textbf{m}_l$) of a Nb ion as a function of $U_{\rm eff}$. The value of $\textbf{m}_n$ is small (ideally to be zero) for the $J_{\rm eff}$=$3/2$ $\Phi_1$ state. When $U_{\rm eff}$ is small, the Hubbard splitting among $\Phi_i$'s ($i$=$1$-$4$) are not sufficient, which reduces the orbital magnetization of occupied state.}
\label{F2}
\end{figure}

\textit{The $J_{\rm eff}$=$3/2$ state}. The crystal structure is relaxed first, which leads to the lattice constants very close to the experimental ones (within $\pm0.8\%$) \cite{Henke:Zk}. To verify the $J_{\rm eff}$ state, the electronic structure is calculated, as shown in Fig.~\ref{F2}(a-d). The $4d$ orbitals of Nb are split into triplet $t_{\rm 2g}$ and doublet $e_{\rm g}$ orbitals by the octaheral crystal field. The Jahn-Teller distortion of NbCl$_6$ octahedra, i.e. the elongation long the $c$-axis, further splits the $t_{\rm 2g}$ and $e_{\rm g}$ orbitals. Then the $t_{\rm 2g}$ triplets become one high-lying $d_{xy}$ and two low-lying $d_{yz}$ and $d_{xz}$. However, such Jahn-Teller splitting is weak, since the length difference of Nb-Cl bonds along the $c$-axis and within the $ab$-plane is very small ($2.431$ \AA{} {\it vs} $2.409$ \AA) and low charge of Cl$^-$ (half of O$^{2-}$).

For Nb$^{4+}$ cation, there is only one $4d$ electron and the spin-polarized GGA calculation yields a metallic result as shown in Fig.~\ref{F2}(a). For occupied states, the $d_{xy}$ contribution is slightly lower than those from $d_{yz}$ and $d_{xz}$, due to the weak Jahn-Teller splitting, when SOC is not included. Without SOC, the addition of Hubbard interaction $U$ will further split the lower Hubbard band and upper Hubbard band of $t_{\rm 2g}$ orbitals (also the orbital disproportion between $d_{yz}$/$d_{xz}$ and $d_{xy}$) as shown in Fig.~\ref{F2}(b). A pseudo gap seems to form at the Fermi level with increasing $U$, but the band gap is not opened till $U_{\rm eff}$=$2$ eV. By considering the SOC but without $U$, the $t_{\rm 2g}$ orbitals is split to the low-lying $J_{\rm eff}$=$3/2$ quartets and high-lying $J_{\rm eff}$=$1/2$ doublets, as expected. The $J_{\rm eff}$=$3/2$ is partially occupied and the system remains metallic, as shown in Fig.~\ref{F2}(c). Then a band gap can be opened by a moderate $U$, rendering a SOC Mott insulator, as shown in Fig.~\ref{F2}(d). The underlying physical mechanism for such SOC Mott insulator is very similar to the $J_{\rm eff}$=$1/2$ state for $d^5$ systems, as demonstrated in Sr$_2$IrO$_4$ \cite{Kim:Prl08}.

Noting although $J_{\rm eff}$=$3/2$ state was claimed in $M_2$TaCl$_6$ experimentally \cite{Ishikawa:Prb19}, its Mottness remains unstudied. The more expanding spatial distribution and weaker Hubbard correlation of $5d$ orbitals may lead to a metallic state like Fig.~\ref{F2}(c), which allows the certain mixture among $J_{\rm eff}$ states. In this sense, the $4d^1$ system studied here may be the best candidate to get the $J_{\rm eff}$=$3/2$ Mott state, as a result of reduced kinetic energy and subtle balance between SOC and Hubbard correlation.

\begin{table}
\caption{The real and imaginary components of the spin-orbital-projected wave function of topmost valence band at the $\Gamma$ point ($U_{\rm eff}$=$1$ eV), which is very close to the ideal $\Phi_1$. Due to the choose of Wigner-Seitz sphere of Nb ion during the projection, the amplitude of wavefunction is not ideally normalized. Even though, the ratios among orbitals are very close to the expected ones for $\Phi_1\sim|d_{yz}\uparrow>+i|d_{xz}\uparrow>$.}
\centering
\begin{tabular*}{0.46\textwidth}{@{\extracolsep{\fill}}lcccccc}
\hline \hline
Bases & $d_{xy}\uparrow$ & $d_{yz}\uparrow$ & $d_{xz}\uparrow$ & $d_{xy}\downarrow$ & $d_{yz}\downarrow$ & $d_{xz}\downarrow$ \\
\hline
real & $0$ & $0.635$ & $0.012$ & $0$ & $0$ & $0$ \\
imaginary & $0$ & $-0.020$ & $0.635$ & $0$ & $0$ & $0$ \\
\hline \hline
\end{tabular*}
\label{table1}
\end{table}

In the atomic limit, the wave functions of ideal $J_{\rm eff}$=$3/2$ quartets are: $\Phi_1$=($|d_{yz}\uparrow>+i|d_{xz}\uparrow>$)/$\sqrt{2}$, $\Phi_2$=($|d_{yz}\downarrow>-i|d_{xz}\downarrow>$)/$\sqrt{2}$, $\Phi_3$=($|d_{yz}\downarrow>+i|d_{xz}\downarrow>-2|d_{xy}\uparrow>$)/$\sqrt{6}$, and $\Phi_4$=($|d_{yz}\uparrow>-i|d_{xz}\uparrow>+2|d_{xy}\downarrow>$)/$\sqrt{6}$, where $\uparrow$/$\downarrow$ denote the spin up/down. Here, due to the weak Jahn-Teller splitting (i.e. the on-site energy of $d_{xy}$ is slightly higher than that of $d_{xz}$/$d_{yz}$), $\Phi_1$ and $\Phi_2$ will be slightly lower in energy than $\Phi_3$ and $\Phi_4$ (In fact, $\Phi_3$ and $\Phi_4$ will be distorted from their ideal limits). Then the spin-up occupied state in the SOC Mott state should be $\Phi_1$, which is confirmed by the wave function extracted from DFT calculation (Table~\ref{table1}). The real and imaginary parts of DFT wave function at the $\Gamma$ point are indeed in the form of $|d_{yz}\uparrow>+i|d_{xz}\uparrow>$, a decisive fringerprint of spin-orbit entangled $\Phi_1$ state.

For the $\Phi_1$ state, the $z$-component of spin moment is $<$$\Phi_1|S^z|\Phi_1$$>$=$1/2$, while the $z$-component of orbital moment is $<$$\Phi_1|L^z|\Phi_1$$>$=$-1$. Considering the ratio of Lande factors ($g_{\rm spin}$=$2$ and $g_{\rm orbit}$=$1$), the magnetization from spin moment and orbital moment should be fully compensated, leading to a ``recessive" magnetic state, i.e. ordered zero magnetic moments. Our spin polarized GGA+$U$+SOC calculation indeed finds that the magnetic moments contributed by spin and orbital are very close in magnitude with opposite signs in the SOC Mott state, e.g. the net magnetic moment is about $0.18$ $\mu_{\rm B}$/Nb when $U_{\rm eff}$=$1$ eV, as shown in Fig.~\ref{F2}(e). Noting that the compensation between spin and orbital magnetizations is a characteristic of $J_{\rm eff}$=$3/2$ state, while in the $J_{\rm eff}$=$1/2$ state the magnetic moments from spin and orbital are parallel \cite{Kim:Prl08}.

\begin{figure}
\centering
\includegraphics[width=0.46\textwidth]{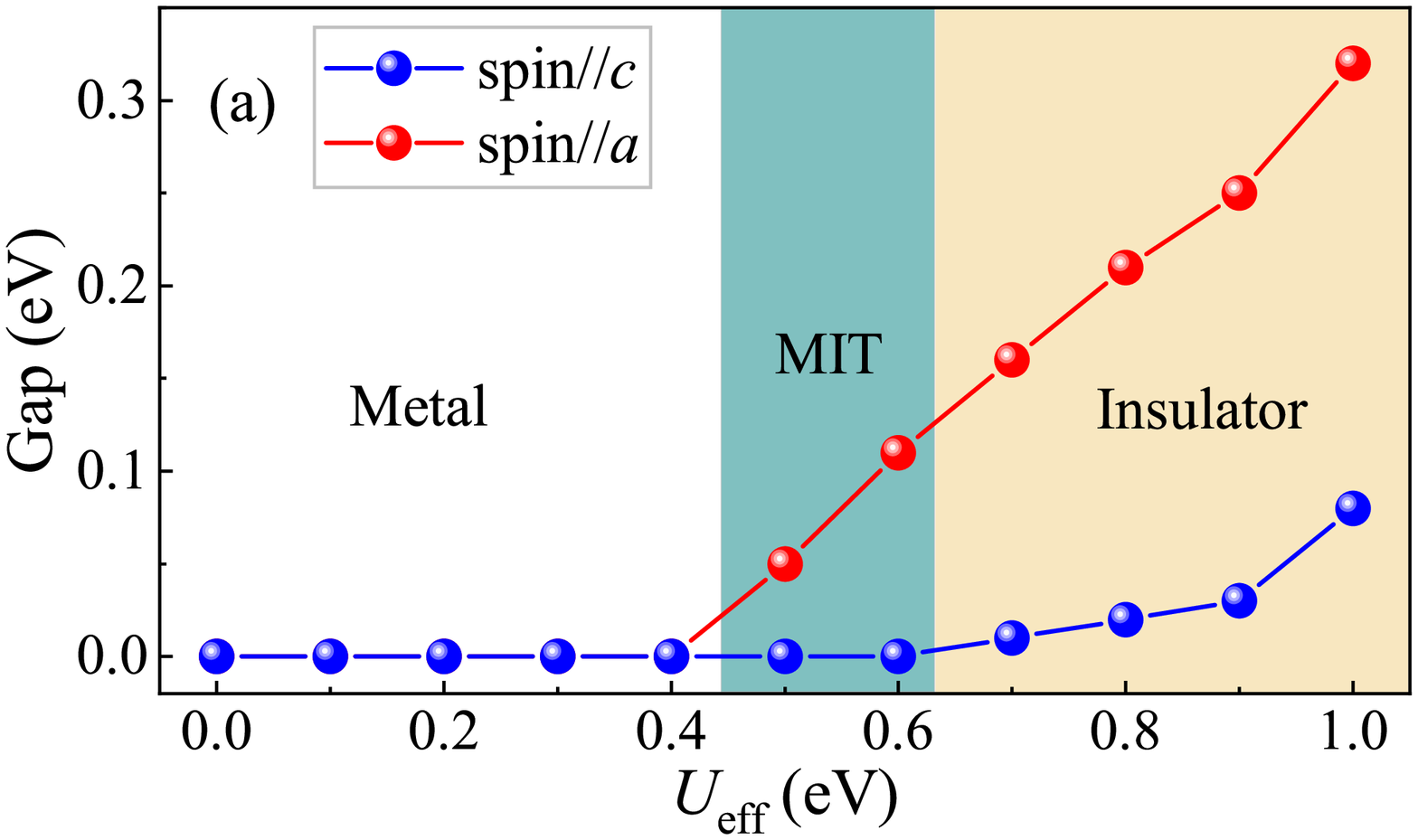}
\includegraphics[width=0.46\textwidth]{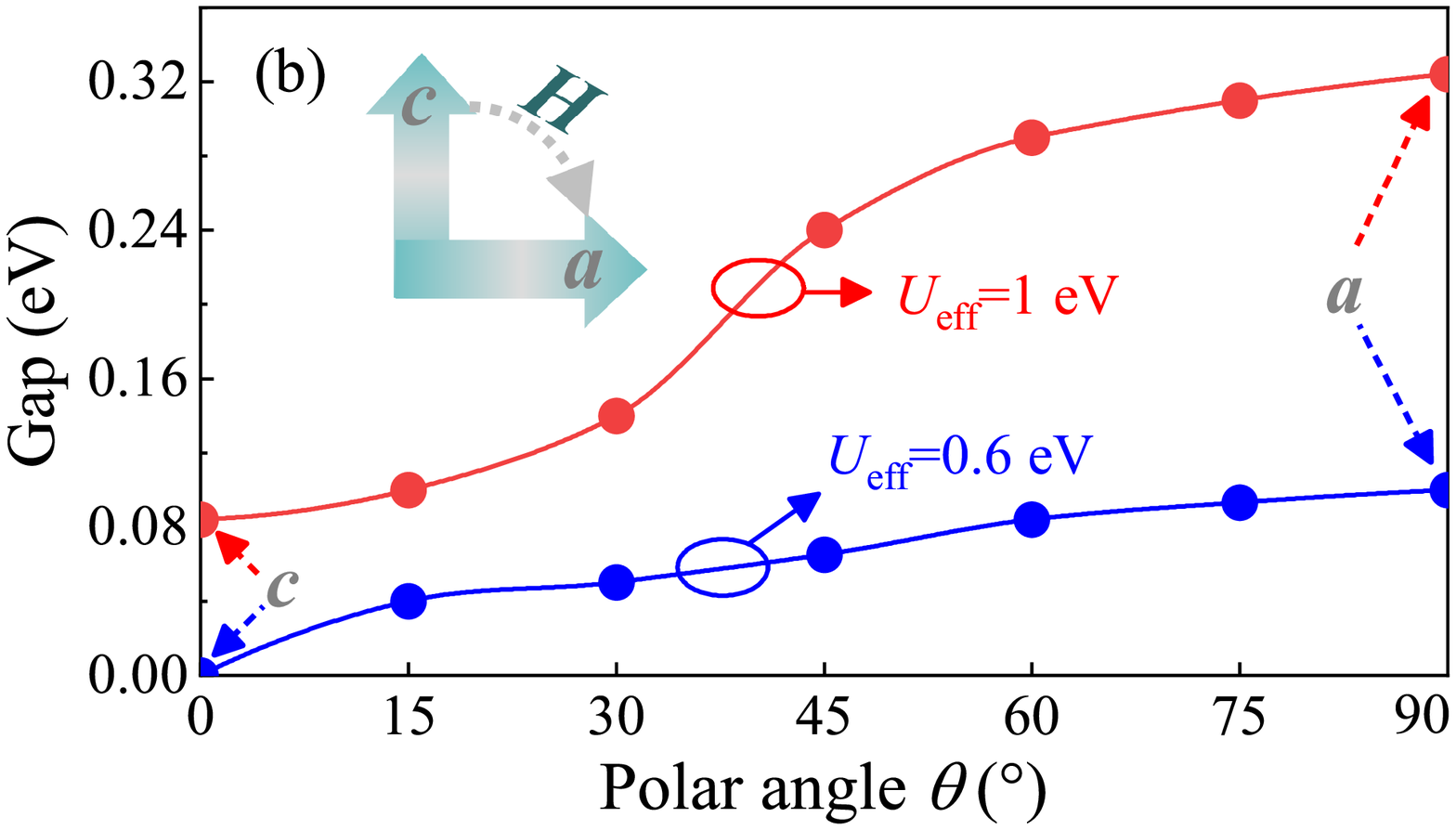}
\includegraphics[width=0.46\textwidth]{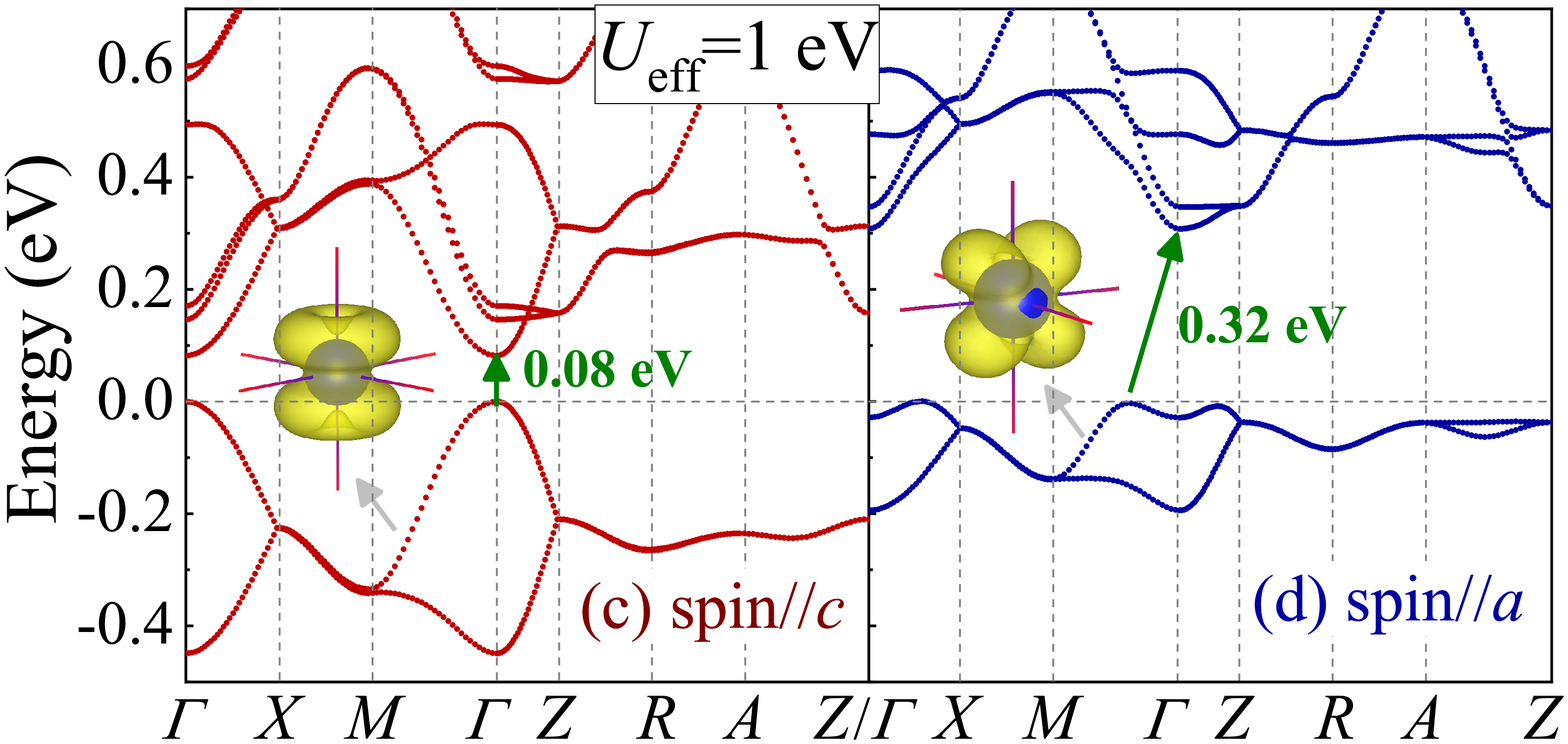}
\caption{(a) The band gap with SOC as a function of $U_{\rm eff}$ for spin$//c$ and spin$//a$ respectively. The metal-insulator transition (MIT) can occur by rotating the spin orientation in the middle $U_{\rm eff}$ region. (b) The band gap with SOC as a function of polar angle $\theta$ for $U_{\rm eff}$=$1$ eV and $U_{\rm eff}$=$0.6$ eV, respectively. Here, $\theta=0^{\circ}$ and $\theta=90^{\circ}$ stand for spin//$c$ and spin//$a$, respectively. Inset: sketch of spin rotation driven by magnetic field. (c-d) The band structures with SOC for $U_{\rm eff}$=$1$ eV: (c) spin$//c$; (d) spin$//a$. Inserts: the corresponding occupied orbital shapes.}
\label{F3}
\end{figure}

\textit{Spin orientation effects}. As discussed before, the tetragonality leads to the ground state with $\Phi_1$ (or $\Phi_2$) occupation, and the wave function $\Phi_1$ (or $\Phi_2$) leads to the magnetocrystalline easy axis along the $c$ axis, although the SOC itself does not break the spatial rotation symmetry of $t_{\rm 2g}$ orbitals.

Starting from this ground state, it is interesting to tune the orbital occupations (and then other physical properties) by rotating the spin axis, which can be realized in real materials via magnetic field since small residual magnetization remains. As shown in Fig.~\ref{F3}(a), a reasonable parameter space of $U_{\rm eff}$ is scanned to reveal the changes of band gap. For different spin orientations, i.e. along the $c$ axis (spin$//c$) \textit{vs} the $a$ axis (spin$//a$), the system evolves asynchronously as a function of Hubbard $U_{\rm eff}$. Especially in the middle region, e.g. $U_{\rm eff}$=$0.5$-$0.6$ eV, the system is metallic when spin$//c$ but insulating when spin$//a$, i.e. a quite exotic spin-orientation determined metal-insulator transition.

Even in the large $U_{\rm eff}$ region where the system is completely an insulator, the change of band gap is quite significant, e.g. the on-off ratio reaches $\sim400\%$ ($0.08$ eV for spin$//c$ and $0.32$ eV spin$//a$) at $U_{\rm eff}$=$1$ eV, which will lead to a much stronger effect comparing with the anisotropic magnetoresisitive effect observed in Sr$_2$IrO$_4$ \cite{Lu:Afm18,Wang:Nc19}. The modulation of band gap is continuous as a function of polar angle [Fig.~\ref{F3}(b)]. Also, the type of band gap changes from the direct type (for spin$//c$) to indirect type (for spin$//a$), as shown in Fig.~\ref{F3}(c-d). Such spin orientation-dependent band gap is also confirmed in the HSE calculations, although the gap becomes larger \cite{Supp}.

When spin rotates from the $c$-axis to $a$-axis, the bandwidth of occupied state is also significantly reduced, leading to heavier hole carriers. Meanwhile, the electronic cloud of occupied $4d$ orbital changes from the mostly-$J_{\rm eff}$=$3/2$ one to a mostly-$S$=$1/2$ one, as visualized in the insets of Fig.~\ref{F3}(c-d). Such transition is due to the Jahn-Teller distortion driven non-ideal $\Phi_3$ and $\Phi_4$ as discussed before. In other words, in such a tetragonal octahedron, $\Phi_1$ and $\Phi_2$ are more closer to the $J_{\rm eff}$=$3/2$ ideal limit, while $\Phi_3$ and $\Phi_4$ are more distorted and thus non-ideal. Luckily, this broken degeneration provides a function to tune its electronic structure via spin rotation.

\begin{figure}
\centering
\includegraphics[width=0.46\textwidth]{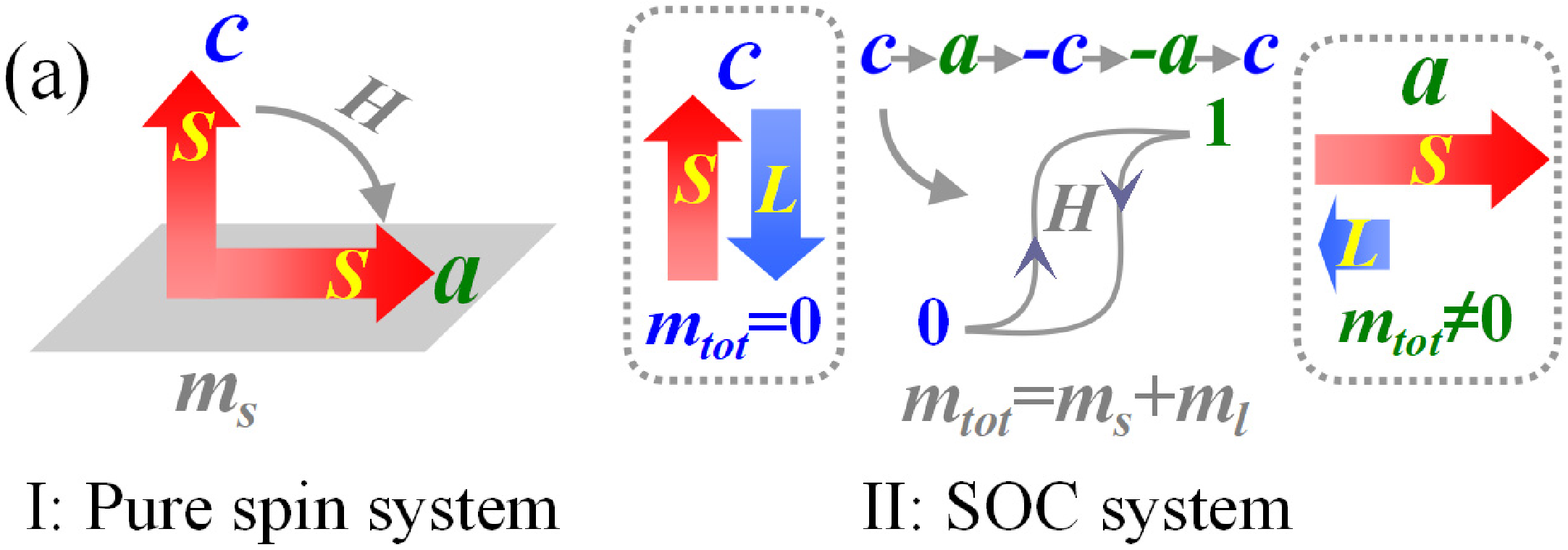}
\includegraphics[width=0.46\textwidth]{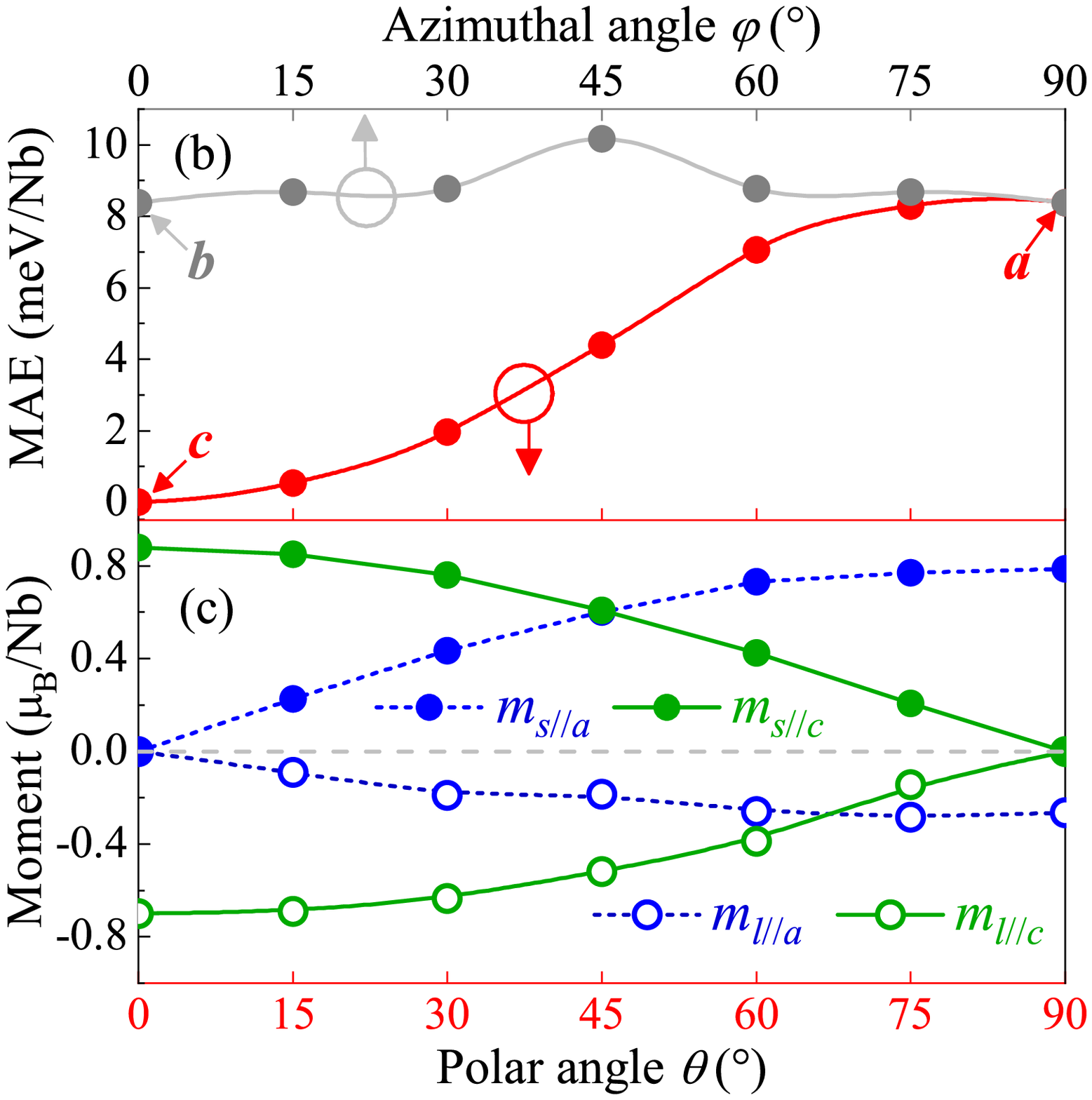}
\caption{(a) Schematic function of spin orientation control of magnetism. For pure spin systems (left), when a spin rotates, the amplitude of its magnetic moment ($m_s$) is a constant and the electronic structure will be unchanged. For those SOC $J_{\rm eff}=3/2$ systems (right), the amplitude of magnetic moment $m_{tot}$ (contributed by both spin $m_s$ and orbital $m_l$ components) can be tuned accompanying the rotation of spin orientation, which is originated from the change of electronic structure. (b) Magnetocrystalline anisotropy. The energy for spin//$c$ is taken as a reference. Upper axis: rotation of azimuthal angle. Lower axis: rotation of polar angle. (c) The $a$/$c$ components of local spin and orbital moments (i.e. $\textbf{m}_s$ and $\textbf{m}_l$) as a function of polar angle at $U_{\rm eff}$=$1$ eV.}
\label{F4}
\end{figure}

Besides the significant tuning of band structure, the magnetic moment can also be strongly modulated by spin orientation, as sketched in Fig.~\ref{F4}(a). Figure~\ref{F4}(b) shows the energies as function of spin rotation angles. The in-plane spin rotation from $a$ axis to $b$ axis can only lead to tiny energy fluctuation ($<2$ meV/Nb), which is reasonable considering the tetragonality. Such tiny fluctuation comes from the octahedral rotation as shown in Fig.~\ref{F1}(c). In contrast, the spin rotation from $c$ axis to $a$ axis needs to overcome a large magnetocrystalline anisotropy energy (MAE) $\sim8.4$ meV/Nb at $U_{\rm eff}$=$1$ eV.

As shown in Fig.~\ref{F4}(c), with increasing polar angle of spin, the local orbital moment $\textbf{m}_l$ also rotates synchronously, but the magnitude of $\textbf{m}_l$ is seriously reduced when spin$//a$. Then the compensation between $\textbf{m}_l$ and $\textbf{m}_s$ for spin$//a$ is partially suppressed, leading to a larger local magnetic moment ($\textbf{m}_s+\textbf{m}_l$) $\sim0.5$ $\mu_{\rm B}$/Nb, comparing with $0.18$ $\mu_{\rm B}$/Nb for spin$//c$. In other words, the characteristic of $S$=$1/2$ state appears over the original $J_{\rm eff}$=$3/2$ state, in agreement with the aforementioned electronic structures.

Finally, it should be noted that although our work only focuses on a special material K$_2$NbCl$_6$, the physical mechanism revealed here is generally applicable for other hexachloro niobates, and even more $4d$/$5d$ transition metal halides and oxides.

\textit{Summary}. A new quantum material K$_2$NbCl$_6$, as an ideal platform for $J_{\rm eff}$=$3/2$ SOC Mott state, was theoretically investigated. The main characteristics of $J_{\rm eff}$=$3/2$ state, including the collaborative $U$-SOC effect, complex wave function, mostly-compensated local magnetic moment, have been unambiguously revealed. More interestingly, the strong tuning of electronic structure as well as the local magnetic moment was theoretically realized by rotating spin orientation, which will lead to giant anisotropic magnetoresisitive effect and even metal-insulator transition. Our results not only extend the scope of new $J_{\rm eff}$ materials, but also suggest new efficient routes to utilize these quantum materials.

\begin{acknowledgments}
The authors are grateful to Prof. Hanghui Chen, Prof. Jiangang He and Dr. Xuezeng Lu for illuminating discussions. This work was supported by the National Natural Science Foundation of China (Grant Nos. 11804168, 11834002, 11674055, and U1732126), the Natural Science Foundation of Jiangsu Province (Grant No. BK20180736).
\end{acknowledgments}

\bibliographystyle{apsrev4-2}
\bibliography{./ref}
\end{document}